\documentclass[12pt]{article}
\usepackage{graphics,cite,amssymb,epsfig,float}
\usepackage[usenames,dvips]{color}

\begin{document} 
 
\newcommand{\lsim}{\raisebox{-0.13cm}{~\shortstack{$<$ \\[-0.07cm] $\sim$}}~} 
\newcommand{\gsim}{\raisebox{-0.13cm}{~\shortstack{$>$ \\[-0.07cm] $\sim$}}~} 
\newcommand{\ra}{\rightarrow} 
\newcommand{\lra}{\longrightarrow} 
\newcommand{\ee}{e^+e^-} 
\newcommand{\gam}{\gamma \gamma} 
\newcommand{\nn}{\noindent} 
\newcommand{\non}{\nonumber} 
\newcommand{\beq}{\begin{eqnarray}} 
\newcommand{\eeq}{\end{eqnarray}} 
\newcommand{\s}{\smallskip} 
\newcommand{\tb}{\tan\beta}  

\newcommand{\nmtools}{{\sc NMSSMTools}}
\def\NPB{Nucl. Phys. B} 
\def\PLB{Phys. Lett. B} 
\def\PRL{Phys. Rev. Lett.} 
\def\PRD{Phys. Rev. D} 
\def\ZPC{Z. Phys. C} 
\baselineskip=15.5pt 
 
\rightline{LPT--Orsay 07-135}
\rightline{SHEP-07-46}
\rightline{BONN-TH-2008-01}
\rightline{PTA/07-062}

\vspace{0.7cm} 
 
\begin{center} 

{\Large {\bf Benchmark scenarios for the NMSSM}} 
 
\vspace{0.7cm}

{\sc A. Djouadi$^{1,2,3}$, M. Drees$^3$, U. Ellwanger$^1$, R.Godbole$^4$,   C.
Hugonie$^5$,  \\ S.F. King$^2$,  S. Lehti$^6$,  S. Moretti$^{1,2}$,  A.
Nikitenko$^{7}$, I. Rottl\"ander$^{3}$,  \\ M. Schumacher$^{8}$,  A. M.
Teixeira$^1$} 

\vspace{0.7cm}

$^1$ Laboratoire de Physique Th\'eorique, Universit\'e Paris--Sud, 
F--91405 Orsay Cedex, France. 

$^2$ School of Physics and Astronomy, University of Southampton, Highfield, 
SO17 1BJ, UK. 

$^3$ Physikalisches Institut, University of Bonn, Nussallee 12, 53115 Bonn,
Germany.

$^4$ Center for High Energy Physics, Indian Institute of Science, Bangalore 560
012, India.

$^5$ LPTA, Universit\'e de Montpellier II, 34095 Montpellier, France. 

$^6$ Helsinki Institute of Physics, P.O. Box 64, FIN-00014, University of
Helsinki, Finland.

$^{7}$   Physics Department, Imperial College, Prince Consort Road, London SW7
2AZ,  UK \\ (on leave of absence from ITEP, Moscow, Russia).

$^{8}$ Fachbereich Physik, University of Siegen, Walter Flex Str. 3, 57068
Siegen, Germany.

\end{center} 
 
\vspace{0.5cm}  
\begin{abstract}   

We discuss constrained and semi--constrained versions of the next--to--minimal
supersymmetric extension of the Standard Model (NMSSM) in which a singlet Higgs
superfield is added to the two doublet superfields that are present in the 
minimal extension (MSSM). This leads to a richer Higgs and neutralino spectrum
and allows for many interesting phenomena that are not present in the MSSM. In
particular, light Higgs particles  are still allowed by current constraints and 
could appear as decay products of the heavier Higgs states, rendering their
search rather difficult at the LHC.   We propose benchmark scenarios which
address the new phenomenological features, consistent with present constraints
from colliders and with the dark matter relic density, and with (semi--)universal
soft terms at the GUT scale. We present the corresponding spectra for the Higgs
particles, their couplings to gauge bosons and fermions and their most important
decay branching ratios.  A brief survey of the search strategies for these
states at the LHC is given.  

\end{abstract}

\newpage

\subsection*{1. Introduction}  

The next-to-minimal supersymmetric extension of the standard model (NMSSM), in
which the spectrum of the minimal supersymmetric extension (MSSM) is extended by
one singlet superfield, has been discussed since the early days of supersymmetry
(SUSY) model--building~\cite{genNMSSM1,genNMSSM2,genNMSSM3}.  In the last
decade, the NMSSM gained a renewed interest in view of its positive features as
compared  to the widely studied MSSM. Firstly, the NMSSM naturally solves in an
elegant way the so--called $\mu$ problem~\cite{MuProblem} of the MSSM: to have
an acceptable phenome\-nology, a value in the vicinity of the electroweak or
SUSY breaking scale is needed for the supersymmetric Higgs mass parameter $\mu$;
this is automatically achieved in the NMSSM, since the $\mu$-parameter is
dynamically generated when the singlet Higgs field acquires a vacuum expectation
value of the order of the SUSY breaking scale, leading to a fundamentalLagrangian that  contains no dimensionful parameters apart from the soft  SUSY
breaking terms. Secondly, as compared to the MSSM, the NMSSM can induce a richer
phenomenology in the Higgs and neutralino sectors, both in collider and dark
matter (DM) experiments: on the one hand, heavier Higgs states can decay into
lighter ones with sizable rates~\cite{lighthiggs1,NoLoseNMSSM2,
nmhdecay1,FineTuning1,egh1,lighthiggs2,lighthiggsteva1,lighthiggs3,lighthiggsteva2,lighthiggs4,
lighthiggsteva3,cheung} and, on the other hand, a new possibility appears for
achieving the correct cosmological relic density~\cite{relicdens} through the
so-called ``singlino", i.e. the fifth neutralino of the model, which can have
weaker-than-usual couplings to standard model (SM) particles. Thirdly, the NMSSM
needs somewhat less fine  tuning~\cite{FineTuning1,FineTuning2} (although some
fine tuning is still required~\cite{finetuning3}):   the upper limit on the mass
of the lightest CP--even  Higgs particle is larger than in the MSSM, and
therefore more SUSY parameter space survives the bounds imposed by the negative
Higgs boson searches at LEP; furthermore, possible unconventional decays of the
SM--like Higgs scalar allow it to have a relatively small mass, well below the
SM Higgs mass limit of 114 GeV, still consistent with LEP constraints.\s

Given the possibly quite different phenomenology in the Higgs sector  as
compared to the SM and the MSSM \cite{Hreviews}, it is important to  address the
question whether such NMSSM specific scenarios will be probed at the LHC. In
particular, it would be important to extend the so-called ``no-lose theorem" of
the MSSM \cite{nolosemssm}, which states that at least one MSSM Higgs particle
should be observed at the LHC for the planned integrated luminosity, to the
NMSSM  \cite{NoLoseNMSSM1,NoLoseNMSSM2} or try to  define regions of the NMSSM
parameter space where more Higgs states are visible at the LHC than those
available within the MSSM \cite{Moretti:2006sv}.  However, a potential drawback
of the NMSSM, at least in its non constrained versions, is that it leads to a
larger number of input parameters in the Higgs sector to deal with, compared  to
the MSSM. In particular, it is clearly unfeasible to perform multi-dimensional
``continuous"  scans over the free inputs of the NMSSM, especially  if each
sampled point is subject to a complete simulation in order to be as close as
possible to the experimental conditions.\s

An alternative approach, acknowledged by both the theoretical and experimental
communities, is that of resorting to the definition of so-called ``benchmark
points'' (or slopes, or surfaces) in the SUSY parameter
space~\cite{SUSYbenchmarks}. These consist of a few ``discrete"  parameter
configurations of a  given SUSY model, which embody the most
peculiar/representative  phenomenological features of the model's parameter
space. Using discrete points avoids scanning the entire parameter space, 
focusing instead on representative choices that reflect the new interesting 
features of the model, such as  new signals, peculiar mass spectra, etc. A
reduced number of points can then be subject to full experimental investigation,
without loss of substantial theoretical information.\s

While several such benchmark scenarios have been devised for the 
MSSM~\cite{SUSYbenchmarks,MSSMBenchmarks}  and thoroughly studied in both the
collider  and the DM contexts, limited progress has been made so far in this
direction for the case of the NMSSM. In Ref.~\cite{egh1}, an earlier attempt was
made to address the possibility of establishing a no-lose theorem for the NMSSM
Higgs sector at the LHC through benchmark points (in a non-universal low energy
setup in which the SUSY particles are very heavy). However, the corresponding
spectrum was not consistent with DM constraints (in particular on the lightest 
SUSY particle, LSP), which were not yet established
for the NMSSM at the time. In addition, many of the points discussed in
Ref.~\cite{egh1} have become ruled out due to the new lower value of the top
quark mass, and to more stringent constraints  from collider searches and
precision measurements. Also the tools to calculate the Higgs and SUSY particle
spectra have been upgraded since then~\cite{nmhdecay2,nmssmtools}.\s

In this report, we build on the experience of Ref.~\cite{egh1} and define
benchmark points which fulfill the present collider and cosmological constraints
using the most up-to-date tools to calculate the particle spectra. However, in
contrast to Ref.~\cite{egh1}, we work in the framework of a (semi--)constrained
NMSSM parameter space, henceforth called cNMSSM, where the soft SUSY--breaking
parameters are defined at the Grand Unified (GUT) scale:   on the one hand, this
approach leads to a much more plausible sparticle spectrum and allows one to
relate features of the Higgs sector to properties of the neutralino sector (and
hence of the LSP); on the other hand, this restricted parameter space still
contains the phenomenological features of the NMSSM that are very distinctive
from those of the MSSM, and  is suitable for intensive phenomenological
investigation by the experimental collaborations. The emphasis will primarily be
on the different possible scenarios within the Higgs sector and the implication
for Higgs searches at the LHC. However, we will also comment on the possible
implications of these benchmark points for the cosmological relic density of the
lightest neutralino DM candidate. Finally, we will describe the tools used to
define such benchmark scenarios.\s

The report is organized as follows. In the next section, we define the NMSSM and
its particle content with some emphasis on  constrained scenarios defined in
terms of soft terms specified at the  GUT scale, discuss the tools that allow to
calculate the particle spectrum, and the various constraints that should be
imposed on the latter. In section~3, we propose five benchmark points which lead
to Higgs sectors that are different from those of the MSSM and discuss their
main features. In section~4, we outline the possible search strategies at the
LHC for the  Higgs particles in these scenarios. A brief outlook is given in 
section~5. 

\subsection*{2. The NMSSM and its particle spectrum}  
 
\subsubsection*{2.1 The unconstrained NMSSM}

In this paper, we confine ourselves to the NMSSM with a scale invariant
superpotential. Alternative generalizations of the MSSM -- known as the minimal
non-minimal supersymmetric SM (MNSSM), new minimally-extended supersymmetric SM
or nearly-minimal supersymmetric SM (nMSSM) or with additional U(1)' gauge
symmetries -- exist \cite{other-non-minimal}, but these will not be considered
here, nor the case of explicit CP violation \cite{cpvnmssm}. The scale
invariant  superpotential of the NMSSM is given, in terms of (hatted)
superfields, by
\beq
{\cal W} = \lambda \widehat{S} \widehat{H}_u \widehat{H}_d +
\frac{\kappa}{3} \, \widehat{S}^3 + h_t
\widehat{Q}\widehat{H}_u\widehat{t}_R^c - h_b \widehat{Q}
\widehat{H}_d\widehat{b}_R^c  - h_\tau \widehat{L} \widehat{H}_d
\widehat{\tau}_R^c 
\label{supot}
\eeq
in which only the third generation fermions have been included
(with possible neutrino Yukawa couplings have been set to zero), and $\widehat
Q,  \widehat L$ stand for superfields associated with the $(t,b)$ and
$(\tau,\nu_\tau)$  SU(2) doublets. The first two terms substitute the $\mu
\widehat H_u  \widehat H_d$ term in the MSSM superpotential, while the three
last terms  are the usual generalization of the Yukawa interactions. The soft
SUSY breaking terms consist of  the scalar mass terms for the Higgs and sfermion
scalar fields which, in terms of the  fields corresponding to the complex scalar
components of the superfields, are given by
\beq
 -{\cal L}_\mathrm{mass} &=& 
 m_{H_u}^2 | H_u |^2 + m_{H_d}^2 | H_d|^2 + m_{S}^2| S |^2 \non \\
 &+& m_{\tilde Q}^2|{\tilde Q}^2| + m_{\tilde t_R}^2 |{\tilde t}_R^2|
 +  m_{\tilde b_R}^2|{\tilde b}_R^2| +m_{\tilde L}^2|{\tilde L}^2| +
 m_{\tilde  \tau_R}^2|{\tilde \tau}_R^2|\; ,
\eeq
and the trilinear interactions between the sfermion and Higgs fields, 
\beq
-{\cal L}_\mathrm{tril}=  \lambda A_\lambda H_u H_d S + \frac{1}{3}
\kappa  A_\kappa S^3 + h_t A_t \tilde Q H_u \tilde t_R^c - h_b A_b
\tilde Q H_d \tilde b_R^c - h_\tau A_\tau \tilde L H_d \tilde \tau_R^c
+ \mathrm{h.c.}\;.
\eeq
In an unconstrained NMSSM  with non--universal soft terms at the GUT scale, the
three SUSY breaking masses squared for $H_u$, $H_d$ and $S$ appearing in ${\cal
L}_\mathrm{mass}$ can be expressed in terms of their vevs  through the three
minimization conditions of the scalar potential.  Thus, in contrast to the MSSM
(where one has only two free parameters at the tree level, generally chosen to
be the ratio of Higgs vacuum expectation values (vevs) $\tan\beta$ and the mass
of the pseudoscalar Higgs boson), the Higgs sector of the NMSSM is described by
the six parameters
\beq
\lambda\ , \ \kappa\ , \ A_{\lambda} \ , \ A_{\kappa}, \ 
\tan \beta =\ \langle H_u \rangle / \langle H_d \rangle \ \mathrm{and}
\ \mu_\mathrm{eff} = \lambda \langle S \rangle\; .
\eeq
One can choose sign conventions such that the parameters 
$\lambda$ and $\tan\beta$ are positive, while the parameters $\kappa$, 
$A_\lambda$, $A_{\kappa}$ and $\mu_{\mathrm{eff}}$ can have both signs. \s

In addition to the above parameters of the Higgs sector, one needs to specify
the soft SUSY breaking mass terms in eq.~(2) for the scalars, the trilinear
couplings in eq.~(3) as well as the gaugino soft SUSY breaking mass parameters  
to describe the model completely, 
\beq 
-{\cal L}_\mathrm{gauginos}= \frac{1}{2} \bigg[ M_1 \tilde{B}  
\tilde{B}+M_2 \sum_{a=1}^3 \tilde{W}^a \tilde{W}_a +
M_3 \sum_{a=1}^8 \tilde{G}^a \tilde{G}_a  \ + \ {\rm h.c.} 
\bigg].
\eeq

Clearly, in the limit $\lambda \to 0$ with finite $\mu_{\rm eff}$, the NMSSM
turns into the MSSM with a decoupled singlet sector. Whereas the phenomenology
of the NMSSM for  $\lambda \to 0$ could still differ somewhat from the MSSM in
the case  where the lightest SUSY particle  is the singlino   (and hence
with the possibility of a long lived next-to-lightest SUSY particle
\cite{singlsp}),  we will not consider this situation here. 

\subsubsection*{2.2 The constrained NMSSM}  

As the number of input parameters is rather large, one can attempt to define a
constrained model, hereafter called the cNMSSM, in which the soft
SUSY--breaking  parameters are fixed at the GUT scale, leading to only a handful
of free input parameters to cope with. This approach is motivated by the fact
that in many schemes for SUSY--breaking, the soft SUSY breaking parameters are
predicted to be universal at a very high energy scale. For example,  in the
cMSSM or mSUGRA scenario, one imposes a common gaugino mass $M_{1/2}$, a scalar
mass $m_0$ and a trilinear coupling $A_0$  at $M_{\rm GUT}$, leading to only
four continuous free parameters (together with $\tan \beta$) and the sign of 
$\mu$. The values of the numerous soft SUSY--breaking  parameters at low
energies are then obtained through the renormalization  group evolution (RGE).
Analogously, the cNMSSM allows to describe the entire sparticle spectrum,
including the chargino and neutralino sectors, in terms of a small number  of
free parameters. This is in contrast to the usual procedure where one postulates
universal sfermion masses directly at the weak scale which seems less
plausible since these masses would then be non--universal at the high (GUT)
scale.\s

In analogy to the cMSSM, one can impose unification  of the soft SUSY--breaking
gaugino masses, sfermion and Higgs masses and  trilinear couplings at the scale
$M_{\rm GUT}$:
\beq
M_{1,2,3} \equiv M_{1/2} \ , \ \
m_{\tilde{F}_i} = m_{H_i} \equiv  m_0 \ , \ \
A_{i} \equiv  A_0 \; .
\label{univsoft} 
\eeq

The cMSSM has two additional parameters ($\mu$ and $B$) beyond those in
eq.~(\ref{univsoft}), but the correct value of $M_{Z^0}$ imposes one constraint
(typically used to determine $|\mu|$) and $B$ can be replaced by $\tan\beta$.
Likewise, the fully constrained NMSSM has two additional parameters ($\lambda$
and $\kappa$) beyond the three parameters in eq.~(\ref{univsoft}), and the
correct value of $M_{Z^0}$ imposes one constraint. Hence, a priori the number of
free parameters in the cMSSM and the fully constrained NMSSM is the same.\s

In principle, one could minimize the effective potential of the  cNMSSM with
respect to $H_u$, $H_d$ and $S$, and determine the overall scale of the soft
terms in eq.~(\ref{univsoft}) from the correct value of $M_{Z^0}$: this approach
has been pursued in Ref.~\cite{genNMSSM2}. However, since $\tan\beta$ is then
obtained as output (while the top quark Yukawa coupling $h_t$ is an input), it
becomes very difficult to obtain the correct value for $m_{\rm top}$. Also, the
numerical minimization of the effective potential including radiative
corrections is quite computer-time consuming.\s

Therefore it is much more convenient to employ the analytic form of the three
minimization conditions of the effective potential of the NMSSM: for given
$M_{Z^0}$, $\tan\beta$, $\lambda$ and all soft terms at the weak scale (except
for $m_S^2$), these can be solved for $|\mu_{\rm eff}|$ (or $|\left< S\right>
|$), $\kappa$ and $m_S^2$. As in the cMSSM, the sign of $\mu_{\rm eff}$ can
still be chosen at will; in some sense, the determination of the parameter $B$
through the minimization conditions of the MSSM is replaced by the determination
of $\kappa$ in the NMSSM and the additional minimization w.r.t. $\left<
S\right>$ in the NMSSM leads to the  determination of $m_S^2$.\s

Assuming the weak scale soft scalar masses (apart from $m_S^2$) to arise from a
unified scalar mass $m_0^2$, the determined weak scale value of $m_S^2$ -- when
run up to the GUT scale -- will not coincide with $m_0^2$ in general. One could
confine oneself to regions in parameter space where the difference between
$m_S^2$ and $m_0^2$ is negligibly small, but we found that the phenomenology of
this fully constrained NMSSM hardly differs (at least in the Higgs sector on
which we are focusing here) from the one of the cMSSM once present LEP
constraints are imposed. Hence we will relax the hypothesis of complete
unification of the soft terms in the singlet sector since the singlet could play
a special r\^ole\footnote{An example is when the  singlet is mixed with or
identified with a radion originating from a 5d brane world \cite{radion};
however, here we do not consider a Kaluza-Klein mass scale below $M_{\rm
GUT}$.}, and we will allow for both $m_S^2 \neq m_0^2$ and $A_\kappa \neq A_0$
at $M_{\rm GUT}$. Note that, although $m_S^2$ at $M_{\rm GUT}$ can be negative,
this does {\it not} signal an instability of the potential: the direction
$\left< S \right> \to \infty$ is always protected by a quartic self coupling
$\sim \kappa^2$, which leads to $\left< S \right> \sim M_\mathrm{SUSY}$ at the
minimum of the potential. \s

In addition, for some of the benchmark points (see points P4 and P5 below),
we will also relax the unification hypothesis for $m_{H_u}^2$ and $m_{H_d}^2$,
in analogy to corresponding scenarios within the MSSM (called non-universal
Higgs model or  NUHM \cite{nuhm}).  The hypothesis $A_{\lambda} = A_0$ will also
be relaxed for point 5. Such points in parameter space can have additional
unconventional properties, whose phenomenology should also be investigated.

\subsubsection*{2.3 The Higgs and SUSY spectra}

Following the procedure employed by the routine NMSPEC within NMSSMTools
\cite{nmssmtools}, which calculates the spectra of the Higgs and  SUSY particles
in the NMSSM, a point in the parameter space of the cNMSSM is defined by the
soft SUSY breaking terms at $M_{\rm GUT}$ (except for the parameter $m_S^2$),
$\tan\beta$ at the weak scale, $\lambda$ at the SUSY scale (defined as an
average of the first generation squark masses) and the sign of the parameter
$\mu_{\rm eff}$. The parameters $\kappa$, $m_S^2$ and $|\mu_{\rm eff}|$ are
determined at the SUSY scale in terms of the other parameters through the
minimization equations of the scalar potential.\s

 The RGEs for the gauge and Yukawa couplings have to be integrated from the weak
scale up to $M_{\rm GUT}$ (defined by the unification of the gauge  couplings
$g_1$ and $g_2$), and the RGEs for the soft terms from $M_{\rm GUT}$ down to the
weak scale. Since $\kappa$ and $m_S^2$ are computed at the weak scale, and
threshold effects for the gauge and Yukawa couplings depend on the soft terms at
the SUSY scale, some iterations are necessary in order to satisfy the desired
boundary conditions for the soft terms at the GUT scale, but usually the
procedure converges quite rapidly (at least for not too large values of
$\tan\beta$ or $\lambda$). For the most relevant SM parameters, the strong coupling and the
top/bottom quark masses, we chose $\alpha_s(M_{Z^0}) = 0.1172$, $m_b(m_b)^{
\overline{\rm MS}} = 4.214$ GeV and $m_{\mathrm{top}}^{\mathrm{pole}} = 171.4$
GeV. \s

After RGE running is completed, the gluino, chargino, neutralino  and sfermion
masses are computed including dominant one loop corrections to their pole
masses. The lightest scalar Higgs pole mass is determined to the following
accuracy (in a notation where its tree level mass is proportional to
$\mathcal{O}(g^2)$, where $g$ denotes any of the electroweak gauge couplings):
one loop corrections of  $\mathcal{O}(h_{t,b}^4)$ and $\mathcal{O}(h_{t,b}^2\,
g^2)$ are computed exactly; one loop corrections of the orders $g^4$,
$g^2\lambda^2$, $\lambda^4$, $g^2 \kappa^2$, $\kappa^4$ and $\lambda^2\kappa^2$
include only terms involving large logarithms ${\ln}(M_i^2/M_{Z^0}^2)$, where
$M_i$ are potentially large Higgs or sparticle masses, while two loop
corrections of $\mathcal{O}(h_{t,b}^6)$ and $\mathcal{O}(h_{t,b}^4 \alpha_s)$
include only terms with two powers of large logs. \s

Once the pole masses are known, all Higgs decay branching ratios into SM and
SUSY particles (an adaptation of the decays in the MSSM \cite{hsusy} to the 
NMSSM case)  are determined including dominant (mainly QCD) radiative
corrections \cite{hdecay}, as well as sparticle loops  which contribute to the
couplings of a neutral Higgs  to two photons or gluons \cite{hloop}.\s 

When the spectrum and the couplings of the Higgs and SUSY particles are 
computed, available Tevatron and LEP constraints are checked. The results of the
four LEP collaborations, combined by the LEP Higgs working group, are included
\cite{lep}. More specifically, the following experimental constraints are taken
into account:

\begin{itemize}
\vspace*{-2mm}

\item[(i)]  The masses of the neutralino as well as their couplings to the $Z^0$
boson are compared with the LEP constraints from direct searches and from the
invisible $Z^0$ boson width; \vspace*{-2mm}

\item[(ii)] Direct bounds from LEP and Tevatron on the masses of the charged
particles ($h^\pm$, $ \chi^\pm$, $\tilde q$,~$\tilde l$) and the gluino are
taken into account; \vspace*{-2mm}

\item[(iii)] Constraints on the Higgs production rates from all channels studied
at LEP. These include in particular $Z^0 h^0_i$ production, $h^0_i$ being any of
the  CP--even Higgs particles,  with all possible two body decay modes of
$h^0_i$ (into $b$ quarks, $\tau$ leptons, jets, photons or invisible), and all
possible decay modes of $h^0_i$ of the form $h^0_i \to a^0_j a^0_j$,  $a^0_j$
being any of the  CP--odd Higgs particles, with all possible combinations of
$a^0_j$ decays into $b$ quarks, $c$ quarks, $\tau$ leptons and jets. Also
considered is the associated production mode $e^+e^- \to h^0_i a^0_j$ with,
possibly, $h^0_i \to a^0_j a^0_j$. (In practice, for our purposes, only
combinations of $i=1,2$ with $j=1$ are phenomenologically relevant.)
\vspace*{-2mm}

\end{itemize}

We stress that light Higgs scalars (with $M_{h^0_{1,2}} \lesssim 114$ GeV) can
still be allowed by LEP constraints, if either (i) the $Z^0$--$Z^0$-$h^0_{1,2}$
coupling is heavily suppressed (if, for instance, the state $h^0_i$ is
dominantly a gauge singlet); or (ii) $M_{a^0_1} \lesssim 11 \, {\rm GeV}$ such
that the state $h^0_{1,2}$ decays dominantly into $a^0_1 a^0_1$ states, but the
$b\bar b$ decay of $a^0_1$ is  impossible; constraints from the decays
$h^0_{1,2} \to a^0_1 a^0_1 \to 4\tau$ allow for $M_{h^0_{1,2}}$ down to $\sim
86$~GeV.   It is important to note, however, that LEP constraints are
implemented only for individual processes, and that combinations of different
processes could potentially rule out seemingly viable scenarios.\s

Finally, experimental constraints from B physics \cite{bphys} such as  the
branching ratios of the rare decays BR$(B \to X_s \gamma)$,  BR$(B_s
\to \mu^+ \mu^-)$ and BR$(B^+ \to \tau^+ \nu_\tau)$ and the mass
differences $\Delta M_s$ and $\Delta M_d$, are also implemented; compatibility
of each point in parameter space with the current experimental bounds is
required at the two sigma level.\s

The new features of the NMSSM have an impact on the  properties of the lightest
neutralino as a dark matter candidate \cite{relicdens}. For instance,  given the
presence of a fifth neutralino (singlino), the composition of the annihilating
WIMPs can be significantly different from those within the MSSM in wide regions
of the parameter space. In particular, the LSP can have a large singlino
component, in which case one has new couplings of the LSP to singlet-like Higgs
states whose mass can be substantially lighter than the Higgs states within the
MSSM. In the presence of light $h^0_1$ and $a^0_1$ states, there are new
channels through which neutralino annihilation can occur in the NMSSM:
$Z^0\,h^0_1$, $h^0_1\,h^0_1$, $h^0_1\,a^0_1$ and $a^0_1\,a^0_1$, either via
$s$--channel $Z^0,\,a^0_i,\,h^0_i$ or $t$--channel neutralino exchange
\cite{relicdens}. Among the proposed benchmark points, we will find examples of
these new features.\s

Technically, the relic abundance of the NMSSM dark matter candidate $\chi_1^0$
is evaluated via a link to MicrOMEGAS \cite{micromegas}. All the relevant cross
sections for the lightest neutralino annihilation and co-annihilation are
computed. The density evolution equation is numerically solved and the relic
density of $ \chi^0_1$ is calculated. The result is compared with the ``WMAP"
constraint  $0.094 \, \lesssim \Omega_{\rm CDM} h^2 \lesssim 0.136$ at the
$2\sigma$ level \cite{wmap}. 

\subsection*{3. The benchmark points} 

As already mentioned in the introduction, in view of the upcoming LHC, quite
some work has been dedicated to probe the NMSSM Higgs sector over the recent
years. In the NMSSM, two different types of scenarios have been pointed out,
depending on whether  Higgs-to-Higgs decays are kinematically allowed or
forbidden. Within the first category, where Higgs-to-Higgs decays are
kinematically allowed and can be dominant, there are two possibilities, each
associated with light scalar/pseudo\-scalar Higgs states:\s

(i) The lightest pseudoscalar Higgs boson $a^0_1$ is rather light, $M_{a^0_1}
\lsim$ 40--50 GeV, and  the lightest CP--even Higgs particle $h^0_1$ has enough
phase space for the decay into two pseudo\-scalar Higgs particles,  $h^0_1 \to
a^0_1 a^0_1$, to be kinematically accessible. In this case, the branching ratio
for the decay $h^0_1 \to a^0_1 a^0_1$ is typically very large and this new decay
channel is the dominant one. Concerning the lightest pseudoscalar decays, there
are two further possibilities:  either $M_{a^0_1} \gsim 10$~GeV and the $a^0_1$
boson decays into  a pair of $b$ quarks or a pair of tau leptons (with the
former decay being in general dominant), leading to $h^0_1 \to a^0_1 a^0_1 \to
4\tau, 4b$ and $2\tau 2b$ final states, or $M_{a^0_1} \lsim 10$~GeV and the
dominant decay mode of the $a^0_1$ boson is  into a pair of tau leptons, leading
to $h^0_1 \to a^0_1 a^0_1 \to 4\tau$ final states. In this latter scenario, one
can still distinguish  two situations for the $h^0_1$ boson: its mass is either
close to its theoretical upper limit of 130 GeV (within  the cNMSSM), or to the
lower limit of 90 GeV, still in agreement with constraints from  Higgs boson
searches at LEP2.\s

(ii) The lightest CP--even Higgs boson is relatively light, $M_{h^0_1} \lsim 50$
GeV, and decays into $b\bar b$ pairs. (The situation where $M_{h^0_1} \lsim 10$
GeV, such that the latter channel is kinematically closed and the decay $h^0_1
\to \tau^+ \tau^-$ dominates, is very constrained by LEP data.) In this case,
the next-to-lightest CP even Higgs boson $h^0_2$ is SM--like, and has a mass
below $\sim 150$ GeV. Even so, it can still decay into two $h^0_1$ bosons
leading to the final topologies $h^0_2 \to h^0_1 h^0_1 \to 4\tau$, $2\tau 2b$
and $4b$, the latter  final state being largely dominating. \s

The second category of scenarios, where Higgs-to-Higgs decays are suppressed,
includes regions of the parameter space where all Higgs particles are relatively
light  with masses in the range 90--200 GeV. This opens the possibility of
producing all the five neutral and the charged Higgs bosons at the LHC. This
scenario is similar to the so--called  ``intense coupling regime" of the MSSM
\cite{ICR,intense} (which has been shown to be rather difficult to be fully
covered at the LHC), but with two more neutral Higgs particles. \s

In the context of the general NMSSM (without unification constraints), there
exists also a decoupling regime in which all the Higgs bosons are heavy and
decouple from the spectrum except for one CP--even Higgs particle, whose mass
can be  up to $\sim 140$ GeV \cite{boundnmssm}.  Such decoupling scenarios are
difficult to realize within the cNMSSM and, in any case, they would be  very
difficult to disentangle from an MSSM ``$M_{h}^{\rm max}$''-scenario at the LHC,
since the separation of the two would require a determination of the parameter
$\tan\beta$ which is notoriously difficult to achieve at the LHC.\s

In the following, we propose five benchmark points of the NMSSM  parameter
space, P1 to P5, in which the above discussed scenarios are realized. Each point
is representative of distinctive NMSSM features: points P1 to P3 exemplify
scenarios where $h^0_1$ decays into light pseudoscalar states (decaying, in
turn, into  $b \bar b$ or $\tau^+ \tau^-$ final states), P4 illustrates the
NMSSM possibility of a very light $h^0_1$, while P5 corresponds to the case
where all Higgs bosons are rather light. In all cases, the input parameters as
well as the resulting Higgs masses and the relevant decay information are given
in Table~\ref{table:NMP}. In the subsequent Table~\ref{table:NMP:cosmo}, we
summarize  the most relevant features of each point regarding the cosmological
relic density.\s

\begin{figure}[ht]
\vskip 1.cm
\begin{center}
\includegraphics[scale=0.64]{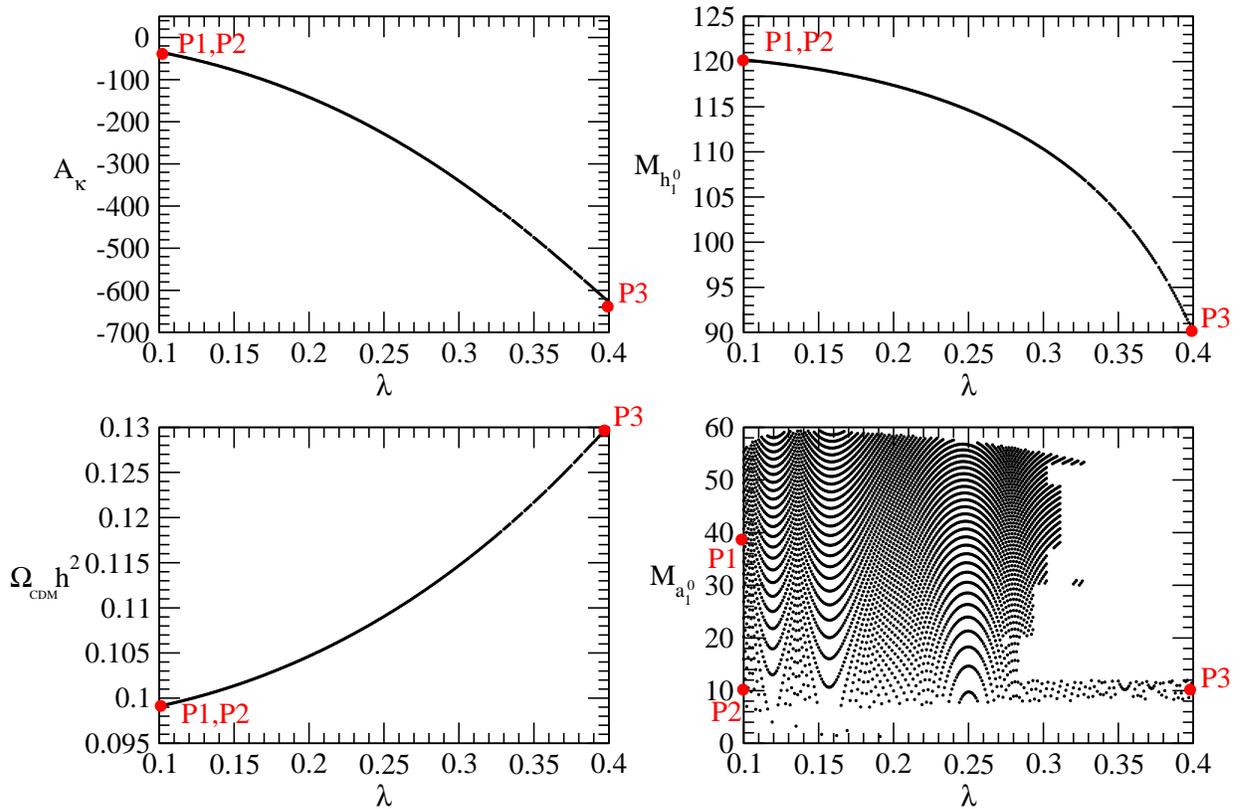}
\vspace*{.6cm}
\caption{Allowed values of $A_\kappa$, $M_{h^0_1}$ and  $M_{a^0_1}$ (in GeV) 
as well as $\Omega_{\rm CDM} h^2$ as a function of 
$\lambda$. We take $m_0 = 174$~GeV, $M_{1/2} = 500$~GeV, $A_0 = -1500$~GeV 
and $\tan\beta = 10$; $m_S^2$ is determined from the electroweak
symmetry breaking conditions.}
\label{1.1f}
\end{center}
\vspace*{-6mm}
\end{figure}

As discussed in Ref.~\cite{nmssmtools}, the scenarios where the pseudoscalar
Higgs boson is light and the decay $h^0_1 \to a^0_1 a^0_1$ is dominant, can be
realized within the cNMSSM with nearly universal soft terms at the GUT scale,
the exception being the parameters $m_S^2$ and $A_\kappa$, which are chosen
independently. A corresponding region in parameter space, containing P1, P2 and
P3, is shown in Fig.~\ref{1.1f}, where we chose $m_0 = 174$~GeV,  $M_{1/2} =
500$~GeV, $A_0 = -1500$~GeV, $\tan\beta = 10$ and the sign  of
$\mu_\mathrm{eff}$ is positive. As a function of the parameter $\lambda$, we
show  the values of $A_\kappa$ that are allowed by LEP constraints, the masses
of the $h_1^0$ and $a_1^0$ bosons (where larger values of $M_{a_1^0}$ correspond to
larger values of the trilinear coupling $|A_\kappa|$), and the dark matter relic
density $\Omega_{\rm CDM} h^2$. The benchmark points P1, P2 and P3 discussed
below are chosen at the upper and lower boundaries of this region  of the cNMSSM
parameter space. The lower limit on the pseudoscalar $a_1^0$ mass,  $M_{a_1^0}
\gsim 8$ GeV, in the last plot of Fig.~1 originates from constraints from B
physics; see Ref.~\cite{bphys} for  details. \s

\begin{figure}[!ht]
\vskip .5cm
\begin{center}
\resizebox{8.cm}{5.5cm}{\includegraphics{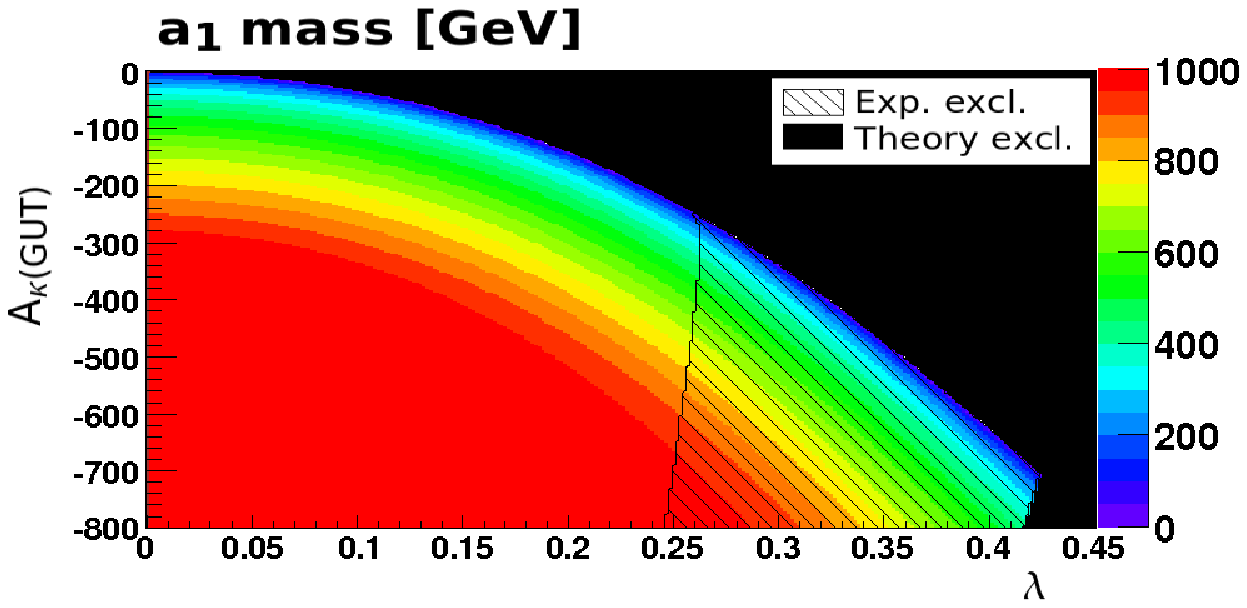}}\hspace*{2mm}
\resizebox{8.cm}{5.5cm}{\includegraphics{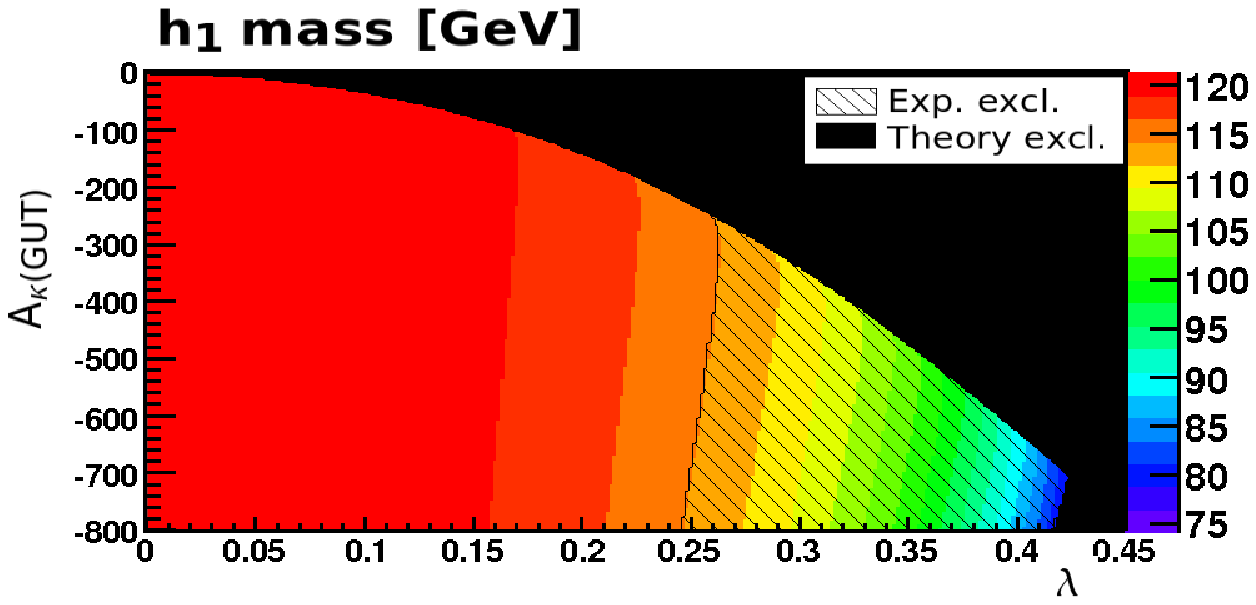}}\vspace*{2mm}
\resizebox{8.cm}{5.5cm}{\includegraphics{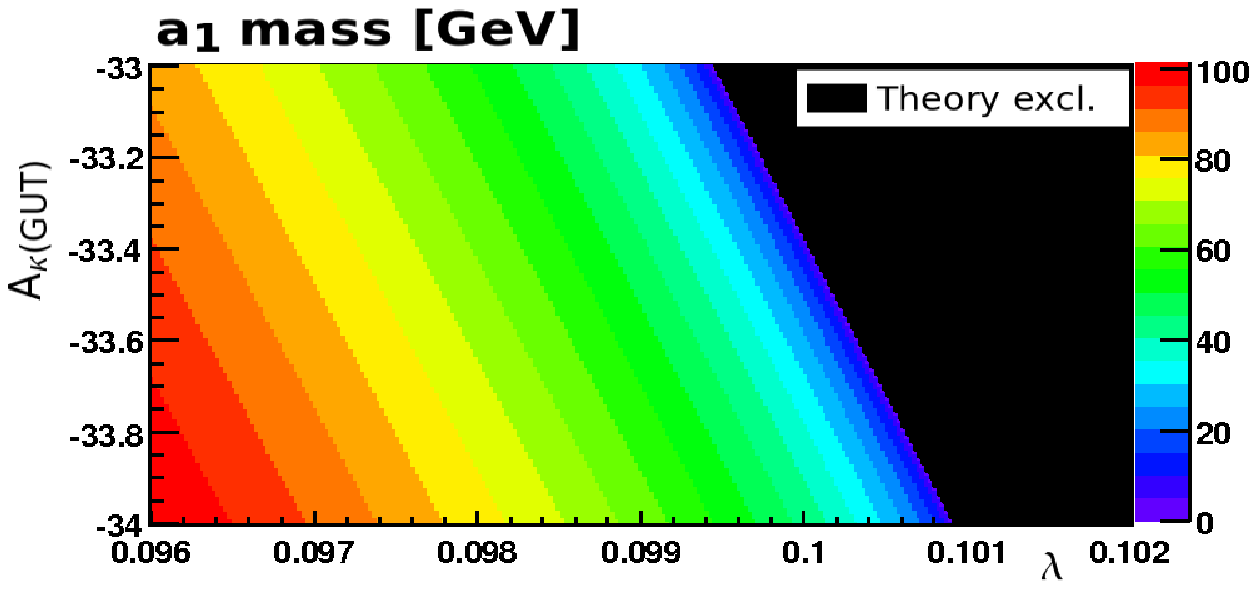}}\hspace*{2mm}
\resizebox{8.cm}{5.5cm}{\includegraphics{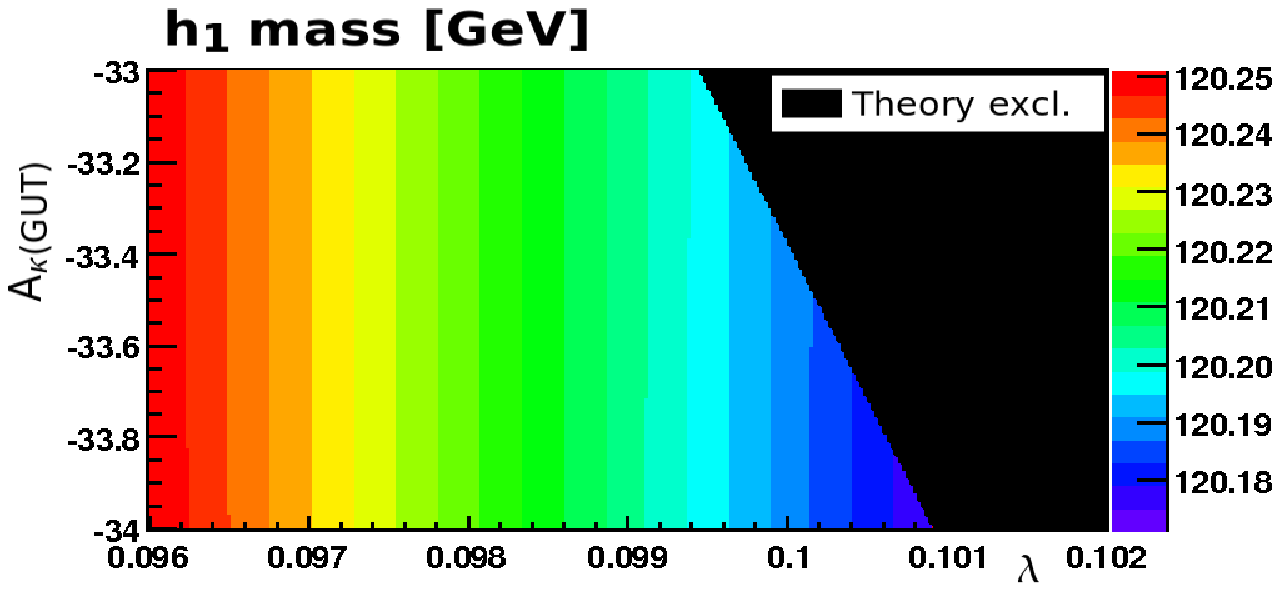}}\vspace*{2mm}
\resizebox{8.cm}{5.5cm}{\includegraphics{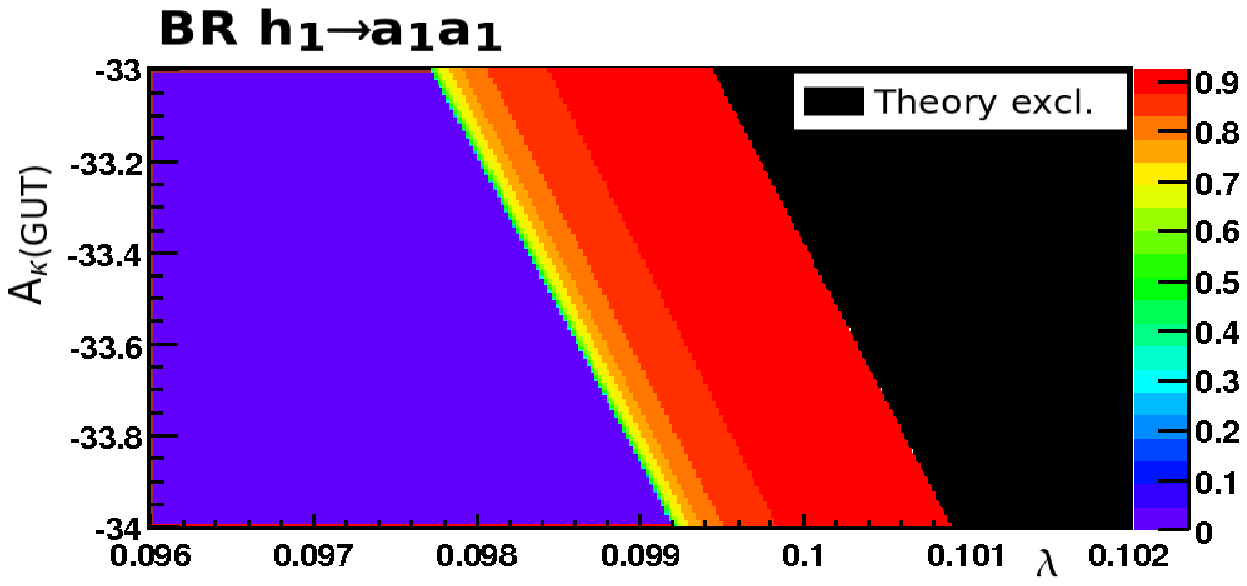}}\hspace*{2mm}
\resizebox{8.cm}{5.5cm}{\includegraphics{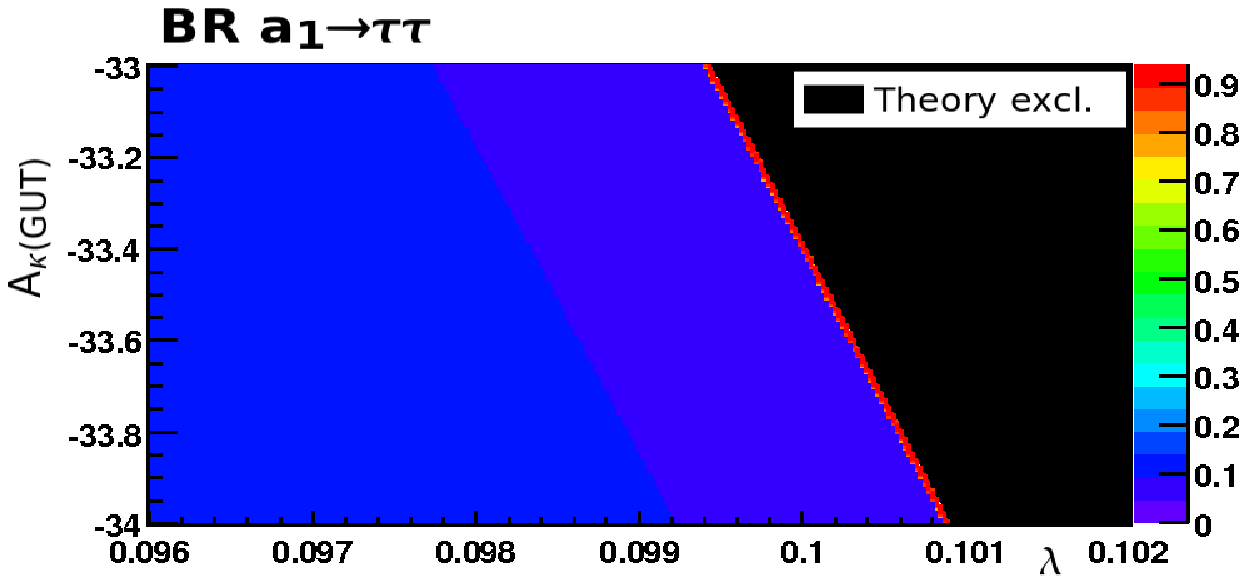}}
\vspace*{0mm}
\caption{Scans in the $[ \lambda, A_\kappa]$ parameter space for the $h^0_1,
a^0_1$ masses, 
and the  branching ratios for the decay modes $h^0_1 \to a^0_1 a^0_1$ and $a^0_1
\to \tau^+ \tau^-$. The two upper frames include P1 to P3 while
the  remaining frames include only P1 and P2; other parameters
are  as in Fig.~1.}
\label{2.1f}
\end{center}
\vspace*{-5mm}
\end{figure}

In Fig.~\ref{2.1f}, we show the obtained  results from a scan in the  [$\lambda,
A_\kappa$] parameter space (with the values of the other parameters fixed to
those used in Fig.~1)  for the masses of the CP--even $h^0_1$  and CP--odd
$a^0_1$ states, as well as of the branching ratios of the decays $h^0_1 \to
a^0_1 a^0_1$ and $a^0_1 \to \tau^+ \tau^-$. In the two upper frames for the
Higgs masses, a scan is performed over the same [$\lambda, A_\kappa$] range as
in Fig.~\ref{1.1f}, which therefore includes  the three points P1, P2 and P3. In
the lower frames for the Higgs masses and   for the branching ratios, a finer
scan zooms on  the [$\lambda, A_\kappa$] range which involves only the two
points P1 and P2. \s

A very light $h^0_1$ state (and, thus, with the decay $h^0_2 \to h^0_1 h^0_1$
being the dominant channel) is represented by P4, which can be obtained once one
relaxes  the universality conditions on the soft SUSY--breaking Higgs mass
terms, $m_{H_d} \neq m_{H_u} \neq m_0$. A scenario as in P5, in which all NMSSM
Higgs bosons are light, is possible if one allows additionally for $A_\lambda
\neq A_0$.\s

For all benchmark points the numerical value of the pseudoscalar Higgs boson
mass $M_{a^0_1}$ is quite sensitive not only to the NMSSM input parameters
(notably to the trilinear coupling $A_\kappa$), but also to the employed SM
parameters and to the precision with which radiative corrections are computed.
Since the numerical values of $M_{a^0_1}$ are phenomenologically much more
important than, for instance, $A_\kappa$ at $M_\mathrm{GUT}$, we consider the
benchmark points to be defined in terms of $M_{a^0_1}$ rather than in terms of
$A_\kappa$ at $M_\mathrm{GUT}$. Next, we summarize the most relevant
phenomenological properties of the benchmark points.\s

In the first two points P1 and P2, the lightest CP--even Higgs boson has a mass
of $M_{h^0_1}\!\simeq\!120$ GeV and is SM--like, as reflected by the
corresponding couplings to gauge bosons $R_1$, top quarks $t_1$ and bottom
quarks $b_1$, which are almost equal to unity, when normalized to the SM Higgs
boson couplings: see Table~\ref{table:NMP}. The lightest CP--odd Higgs boson has
masses of 40.5~GeV and 9.09~GeV,  which are obtained by choosing $A_\kappa$ at
$M_{\rm GUT}$  as  $-33.9$~GeV and $-33.4$~GeV, respectively; the remaining
parameters of the Higgs sector at $M_{\rm SUSY}$, like $\kappa, A_\lambda,
A_\kappa, \mu_{\rm eff}$ and the gaugino mass parameter $M_2$, are given in the
second part of Tab.~\ref{table:NMP}. \s

In both cases P1 and P2, the decay channel $h^0_1 \to a^0_1 a^0_1$ is largely
dominating with a branching rate very close to 90\%, while the decays $h^0_1
\to  b\bar b$ and $\tau^+ \tau^-$ are suppressed by an order of magnitude as
compared to the SM. The most relevant difference between the two scenarios
concerns the mass and decays of the lightest pseudoscalar state. In P1 the
pseudoscalar $a^0_1$ decays into $b$ quarks and $\tau$ leptons, with branching
fractions of $\sim 90\%$ and $\sim 10\%$, respectively. On the other hand, in P2
the pseudoscalar $a^0_1$ with its mass $M_{a^0_1}\!\simeq\!9.09$ GeV decays
dominantly into $\tau^+\tau^-$ pairs, with a rate close to 90\%.\s

For point P3, the same inputs of points P1 and P2 are chosen except for the
$A_\kappa$ and $\lambda$ parameters, which are now varied as to have a lighter
$h^0_1$ state. This again leads to a pseudoscalar Higgs boson  which has
approximately the same mass as in scenario P2, $M_{a^0_1}\!\simeq\!9.96$~GeV,
and which decays almost exclusively into $\tau^+ \tau^-$ final states. The
difference between P3 and P2 is the lightest CP--even Higgs boson $h^0_1$, which
has a mass  $M_{h^0_1}\simeq 90$~GeV, lower than in scenarios P1 and P2. In this
case, and although $h^0_1$ is still SM--like,  exhibiting couplings to gauge
bosons, top and bottom quarks that are very close to those of the SM Higgs
boson, it decays nevertheless almost exclusively into $a^0_1$ pairs, with a rate
close to 100\%.  Another difference between P2 and P3 is that in the former case
the decay mode $h^0_1 \to a^0_1 Z^0$ is kinematically possible but the branching
ratio for this new interesting channel that is not listed in the table is rather
small, BR($h^0_1 \to a^0_1 Z^0) \sim 3\%$.  \s

Note that in all these first three points, the heaviest neutral Higgs particles
$h^0_2, h^0_3$ and $a^0_2$, as well as the charged Higgs states $h^\pm$, all
have masses close to or above 1 TeV. The main decay modes are into $b\bar b$
and  $t\bar t$  for the neutral and $tb$ for the charged states, as $\tb$ is not
too large and the $t\bar t$--Higgs couplings are not very strongly suppressed,
while the branching fractions for the neutral Higgs-to-Higgs decays, in
particular the channels $h^0_2 \to h^0_1 h^0_1$ and $h^0_2 \to  a^0_1 a^0_1$,
are very tiny, not exceeding the permille level.\s

Regarding the properties of the DM candidate, P1, P2 and P3 exhibit a  lightest
neutralino which is bino--like, and whose mass $m_{\chi_1^0}\!\simeq\!208$ GeV.
In all three cases, the correct cosmological density, $\Omega_{\rm CDM}
h^2\simeq 0.1$, is achieved through the co--annihilation  with the $\tilde
\tau_1$ slepton, which has a mass comparable to that of the LSP, $m_{\tilde
\tau_1} \simeq 213, 213$ and 215 GeV, respectively, for P1, P2 and P3. In each
benchmark scenario, the dominant co--annihilation channels are thus $\chi_1^0
\,\tilde \tau_1 \to \gamma \tau$ ($\sim 33\%$) and $\chi_1^0 \,\tilde \tau_1 \to
Z^0 \tau$ ($\sim 10\%$).\s

\begin{table}[!ht]
\caption{Input and output parameters for the five benchmark NMSSM points.}
\vspace*{-5mm}
\label{table:NMP}
\vspace{3mm}
\footnotesize
\begin{center}
\begin{tabular}{|l|r|r|r|r|r|}
\hline
{\bf Point} & P1 & P2 &  P3 & P4 & P5
\\\hline
{\bf GUT/input parameters }
\\\hline
sign($\mu_\mathrm{eff}$)  & + &+ &+ &-- &+
\\\hline
$\tan \beta$  & 10 & 10 & 10 & 2.6& 6
\\\hline
$m_0$ (GeV)  & 174& 174& 174& 775&1500
\\\hline
$M_{1/2}$ (GeV) &500 & 500& 500& 760& 175
\\\hline
$A_0$ & -1500&-1500 & -1500& -2300& -2468
\\\hline
$A_\lambda$ & -1500&-1500 & -1500& -2300& -800
\\\hline
$A_\kappa$ & -33.9& -33.4& -628.56& -1170& 60
\\\hline
NUHM: $M_{H_d}$ (GeV) &-&-&-& 880&-311
\\\hline 
NUHM: $M_{H_u}$ (GeV) &-&-&-& 2195&1910
\\\hline\hline
{\bf Parameters at the SUSY scale } 
\\\hline
$\lambda$ (input parameter) & 0.1& 0.1& 0.4& 0.53&0.016
\\\hline 
$\kappa$ & 0.11 & 0.11& 0.31& 0.12&-0.0029
\\\hline
$A_\lambda$ (GeV) & -982 & -982& -629& -510& 45.8
\\\hline
$A_\kappa$ (GeV) & -1.63& -1.14& -11.4& 220& 60.2
\\\hline
$M_2$ (GeV)& 392 & 392 & 393 & 603 & 140
\\\hline
$\mu_{\rm eff}$ (GeV) & 968 &968 & 936& -193& 303
\\\hline\hline
{\bf CP even Higgs bosons}
\\\hline
$m_{h^0_1}$ (GeV) & 120.2& 120.2& 89.9&32.3 &90.7
\\\hline
$R_1$& 1.00 & 1.00 & 0.998& 0.034&-0.314
\\\hline
$t_1$& 1.00 & 1.00 & 0.999& 0.082&-0.305
\\\hline
$b_1$& 1.018& 1.018& 0.975& -0.291&-0.644
\\\hline
BR($h^0_1 \to b \bar b$) & 0.072& 0.056& $7 \times 10^{-4}$& 0.918&0.895
\\\hline
BR($h^0_1 \to \tau^+ \tau^-$) & 0.008& 0.006& $7 \times 10^{-5}$&
0.073&0.088 \\\hline
BR($h^0_1 \to a^0_1 a^0_1$) & 0.897& 0.921& 0.999& 0.0 & 0.0 
\\\hline\hline
$m_{h^0_2}$ (GeV) & 998 & 998& 964& 123&118
\\\hline
$R_2$& -0.0018& -0.0018& 0.005& 0.999&0.927
\\\hline
$t_2$& -0.102& -0.102& -0.095& 0.994&0.894
\\\hline
$b_2$& 10.00& 10.00& 9.99& 1.038&2.111
\\\hline
BR($h^0_2 \to b \bar b$)& 0.31& 0.31& 0.14& 0.081& 0.87  \\\hline
BR($h^0_2 \to t \bar t$) & 0.11& 0.11& 0.046& 0.0 &0.0 \\\hline
BR($h^0_2 \to a^0_1 Z^0$) & 0.23& 0.23& 0.72& 0.0& 0.0 \\\hline
\hline
$m_{h^0_3}$ (GeV) & 2142& 2142& 1434&547 &174
\\\hline\hline
{\bf CP odd Higgs bosons}
\\\hline\hline
$m_{a^0_1}$ (GeV) & 40.5& 9.1& 9.1 &185&99.6
\\\hline
$t_1^\prime$& 0.0053& 0.0053& 0.0142& 0.0513&-0.00438
\\\hline
$b_1^\prime$& 0.529& 0.528& 1.425& 0.347&-0.158
\\\hline
BR($a^0_1 \to b \bar b$) & 0.91& 0.& 0.& 0.62&0.91
\\\hline
BR($a^0_1 \to \tau^+ \tau^-$) & 0.085& 0.88& 0.88& 0.070&0.090
\\\hline\hline
$m_{a^0_2}$ (GeV) & 1003& 1003 & 996& 546&170
\\\hline\hline
{\bf Charged Higgs boson} 
\\\hline\hline
$m_{h^\pm}$ (GeV) & 1005& 1005& 987& 541&188
\\\hline
\end{tabular}\end{center}
\end{table}

\begin{table}[!ht]
\caption{LSP properties and relic density for the five benchmark NMSSM
  points.}
\label{table:NMP:cosmo}
\footnotesize
\begin{center}
\begin{tabular}{|l|r|r|r|r|r|}
\hline
{\bf Point} & P1 & P2 &  P3 & P4 & P5
\\\hline
{\bf Dark matter } 
\\\hline\hline
LSP ($\chi^0_1$) mass (GeV)  & 208& 208& 208& 101&70.4
\\\hline
$N_{11}$  &  0.999 &  0.999 & 0.999 & -0.039 & 0.977
\\\hline
$N_{12}$  &  -0.008 & -0.008 & -0.009 & 0.043 & -0.098
\\\hline
$N_{13}$  &  0.048 & 0.048 & 0.050 & -0.028 & 0.178
\\\hline
$N_{14}$  &  -0.015 & -0.015 & -0.016 & 0.405 & -0.068
\\\hline
$N_{15}$  &  0 & 0 & 0.003 & 0.912 & -0.003
\\\hline
$\Omega_{\rm CDM} h^2$  & 0.099& 0.099& 0.130& 0.099& 0.105
\\\hline
\end{tabular}\end{center}
\vspace*{-4mm}
\end{table}

The benchmark point P4 corresponds to a scenario in which the CP--even boson
$h^0_1$ is very light, $M_{h^0_1}=32.3$ GeV, and singlet--like,  with a singlet
component above 99\%. The lightest scalar Higgs particle  predominantly decays
into $b\bar b$ pairs, with  BR$(h^0_1 \to b\bar b)=92\%$, and to a smaller
extent into  $\tau$ pairs with BR$(h^0_1 \to \tau^+\tau^-)\simeq 7\%$.  The
CP--even $h^0_2$ boson has a mass of $M_{h^0_2}\!\simeq\!123$ GeV and is
SM--like, with normalized couplings to $W/Z^0$ and $t/b$ states close to unity. 
However, since $M_{h^0_2} >2M_{h^0_1}$, it mostly decays into two $h^0_1$ 
bosons, with BR$(h^0_2 \to h^0_1 h^0_1)\simeq 88\%$, and the dominant SM--like
$b\bar b$ decay mode occurs only at a rate smaller than 10\%. The lightest
CP--odd particle is not very heavy, $M_{a^0_1}=185$ GeV, and decays mostly into
fermion pairs  BR$(a^0_1 \to b\bar b) \sim 61\%$ and  BR$(a^0_1 \to
\tau^+\tau^-) \sim 7\%$. The other dominant decay is the interesting channel
$a^0_1 \to h^0_1 Z^0$ which has a branching ratio of the order of 30\%. Finally,
the heaviest CP even $h^0_3$ and CP--odd $a^0_2$ states and the charged $h^\pm$
particles have masses in the 500 GeV range and will mostly decay, as $\tb$ is
small, into $t\bar t/tb$ final states for the neutral/charged states. All these
features make the  phenomenology of point P4 rather different from that of 
points  P1 to P3 discussed above. \s 

To achieve a correct cosmological relic density, the common sfermion  and
gaugino mass parameter $m_0$ and $M_{1/2}$ at the GUT scale are set close to 1
TeV. At the SUSY scale, one thus finds a higgsino-singlino-like neutralino LSP,
whose mass is $m_{\chi_1^0} \sim 100$ GeV.  LSP annihilation essentially occurs
via two channels: $\chi_1^0 \,\chi_1^0 \to W^+ W^-$ (60\%) and  $\chi_1^0
\,\chi_1^0 \to Z^0 h^0_1$ (20\%), mediated by  $s$--channel $Z^0$ and Higgs
boson exchange.\s  

Finally, point P5 is characterized by having all Higgs particles relatively
light with masses in the range 90 to 190 GeV. Here, the small value for the 
coupling $\lambda$ leads to a small value $\kappa\simeq -0.003$. The three
CP--even Higgs bosons with masses of 91, 118 and 174~GeV, respectively, share
the couplings of the SM  Higgs boson to SM gauge bosons with the dominant
component being taken by the $h^0_2$ state. The reduced couplings of the state 
$h^0_3$ for this point, not given in the Table 1, are $R_3 = -0.205$, $t_3 =
-0.37$ and $b_3 = 5.7$. The pseudoscalar Higgs bosons have masses
$M_{a^0_1}\simeq 100$ GeV and $M_{a^0_2}\simeq 170$ GeV, while the charged Higgs
particle is the heaviest one with a  mass $M_{h^\pm}\simeq 188$ GeV. Here, all
the neutral  Higgs-to--Higgs decays are kinematically forbidden. This is also
the case of neutral Higgs decays into lighter Higgs states with opposite parity
and gauge bosons. The only non--fermionic two--body Higgs decays are thus $h^\pm
\to W h^0_1$ and $h^0_3 \to WW$, but as the involved Higgs--gauge boson
couplings are small, the branching ratios are tiny. \s

In this last point P5, the LSP with a mass $m_{\chi_1^0} \sim 70$~GeV, is a
bino--like neutralino but has a small non--negligible higgsino component. The
correct  cosmological relic density  $\Omega_{\rm CDM} h^2\simeq 0.1$  is
achieved through the exchange of Higgs bosons in the $s$--channel in the
annihilation processes  $\chi_1^0 \,\chi_1^0 \to b \bar b ,\tau^+ \tau^-$ which
occur at a rate  of $\sim 90\%$ and $10\%$, respectively, i.e. comparable to the
Higgs decay branching ratios. \s

More detailed properties of the benchmark points such as more branching ratios
of the Higgs scalars are available on the web site of Ref.~\cite{nmssmtools}.

\subsection*{4. Implications for the LHC}

\subsubsection*{4.1 Dominant production processes}

In the cases discussed here, at least one CP--even Higgs particle $h^0_i$ has
strong enough couplings to massive gauge bosons or top/bottom quarks, i.e. $R_i,
t_i/b_i \sim 1$, to  allow for the production at the LHC in one of the main
channels which are advocated  for the search of the SM Higgs particle 
\cite{Hreviews}:

\begin{itemize}
\vspace*{-2mm}

\item[$i)$] gluon--gluon fusion, $gg \to h^0_i$, occurring through a
loop of heavy  top quarks which couple strongly to the Higgs boson, 
\vspace*{-2mm}

\item[$ii)$] vector boson fusion, $qq\to qq W^*W^*,qqZ^{0*}Z^{0*}\to qq h^0_i$, which
leads to  two forward jets and a centrally decaying Higgs boson,  \vspace*{-2mm}

\item[$iii)$] the Higgs--strahlung processes, $q\bar q' \to Wh^0_i$ and 
$q\bar q \to Z^0h^0_i$, which lead to a Higgs and a massive gauge boson in
the final state, 
\vspace*{-2mm}

\item[$iv)$] associated production with heavy quark pairs $q\bar q/gg \to Q\bar
Q h^0_i$, with $Q=t,b$. 

\vspace*{-2mm} 
\end{itemize}

In the scenarios P1 to P3, this CP--even particle is the $h^0_1$ boson which has
$R_1 \simeq t_1 \simeq b_1 \simeq 1$,  but which decays  predominantly into a
pair of light pseudoscalar Higgs particles, $h^0_1 \to a^0_1 a^0_1$, which
subsequently decay into light fermion pairs, $a^0_1 \to b\bar  b$ and
$\tau^+\tau^-$. In scenario P4, this particle is the $h^0_2$ boson which decays 
most of the time into a pair of $h^0_1$ particles, $h^0_2 \to h^0_1 h^0_1$,
which again   decay into  light fermion pairs. In these four cases, the
backgrounds for  the modes $gg  \to h^0_i \to 4f$ and $qq/gg \to t\bar t h^0_i
\to t\bar t +4 f$ with $f=b,\tau$, will be extremely large and only the vector
boson fusion (owing to the forward jet tagging) and possibly the
Higgs--strahlung (due to the leptons coming from  the decays of the gauge
bosons) can probably be viable at the LHC. Note that in the case of scenario P4,
one can try taking advantage of the  relatively light $a^0_1$ and the large rate
for the interesting and clean decay mode $a^0_1 \to h^0_1 Z^0$, which occurs at
the level of 2\% if  the $Z^0$ boson decays into electrons and muons.  However,
the cross sections for $a^0_1$ production in the $gg \to a^0_1$ and $gg/q\bar q
\to t\bar t a^0_1$ processes, which are the dominant ones for the production of
the MSSM CP--odd Higgs boson  \cite{Hreviews}, are very small as the reduced
$a^0_1 tt$ coupling is very tiny, $t'_1\sim 0.05$.  One would have then to rely
on  associated production of the $a^0_1$ state with a CP--even Higgs boson, such
as $q \bar q \to Z^{0*} \to  a^0_1 h^0_1$, but the cross section  is not 
expected to be very large. \s

In scenario P5, the particle which has couplings to gauge bosons and top quarks 
that are close to those of the SM Higgs boson is the $h^0_2$ boson which decays
into $b\bar b$ and $\tau^+\tau^-$ final states with branching ratios close  to
90\% and 10\%, respectively. Here again, the $gg$ fusion and presumably 
associated production with top quarks cannot be used since the interesting
decays  such as $h^0_2 \to W W^*, Z^0Z^{0*}$ and $\gamma \gamma$ are  suppressed
when compared  to the SM case (by at least a factor of four as $b_2 \sim 2$). 
Thus, in this case, only the channels $qq\to qqh^0_2 \to qq \tau^+ \tau^-$ and
possibly $q\bar q' \to Wh^0_2 \to \ell \nu b \bar b$ seem feasible. The state
$h^0_1$ has still  non--negligible couplings to gauge bosons and top quarks
which lead to cross  sections that are ``only" one order of magnitude smaller
than in the SM. Since here, only the decays $h^0_1 \to b\bar b$ (90\%) and
$\tau^+\tau^-$ (10\%)  are again  relevant, the only processes which can be used
are the vector boson fusion  and Higgs--strahlung processes discussed above.
Even so, one needs a luminosity 10 times larger to have the same event samples
as in the SM. Note that for this point, one can also consider associated
CP--even and CP--odd Higgs production, $q \bar q \to Z^{0*} \to h^0_i a^0_i$ but
the cross sections are still small.  Finally, one should note that for this
point, the value of $\tb$ is moderate and thus,  the charged Higgs coupling to
$tb$ states is the smallest possible, and does not guarantee the detection of
the $h^\pm$ bosons.

\subsubsection*{4.2 Summary of available studies} 

Let us now briefly summarize the few detailed studies (possibly including  Monte
Carlo simulations) of the LHC potential for the NMSSM Higgs sector that have
been performed for scenarios close to the ones discussed here\footnote{ The
prospects  for the Tevatron have been discussed in
Refs.~\cite{lighthiggsteva1,lighthiggsteva2,lighthiggsteva3} and seem quite
dim.}. \s

\underline{In scenarios with a very light pseudoscalar Higgs boson}, preliminary
LHC studies  focused on the $qq\to qq WW,qqZ^0Z^0$ $\to qq h^0_1 \to qq
a^0_1a^0_1$ detection mode, i.e. via vector boson fusion (VBF) with
forward/back\-ward jet tagging \cite{NoLoseNMSSM2}. The hope was that such NMSSM
specific scenarios would be visible, particularly if the lightest CP--odd Higgs
boson mass allowed for abundant $a^0_1a^0_1\to b\bar b\,\tau^+\tau^-$ decays,
with both $\tau$--leptons being detected via their $e,\mu$ leptonic decays. At
high luminosity, the VBF signal may be detectable at the LHC as a bump in the
tail of a rapidly falling mass distribution of the $2\tau\,2j$ system (without
$b$ tagging). However, this procedure relies on the background shape to be
accurately predictable. These analyses were based on Monte Carlo (MC) event
generation via the SUSY routines of the {\tt HERWIG} code \cite{SHERWIG} and the
toy detector simulation {\tt GETJET}. Further analyses based on the {\tt PYTHIA}
generator \cite{Sjostrand:2001yu} and the more adequate ATLAS detector
simulation {\tt ATLFAST} \cite{Baffioni:2004gdr} found that the original
selection procedures may need improvement in order to extract a signal.\s

Also considered  was the Higgs--strahlung process with the $q\bar q'\to W^{*}\to
W^\pm h^0_1 \to W^\pm a^0_1 a^0_1$ signature, exploiting leptonic decays of
gauge bosons with a subleading component from $q\bar q\to Z^{0*}\to Z^0 h^0_1\to
Z^0 a^0_1a^0_1$ and, more marginally, associated production with top quarks,
$q\bar q,gg\to t\bar t h^0_1\to t\bar t a^0_1 a^0_1$ \cite{lighthiggs3}.
Regardless of the decay modes of the pseudoscalar Higgs bosons,  it has been
shown in this parton--level analysis, but in which the efficiency to trigger on
the signal is included, that at least the Higgs--strahlung process $q \bar q \to
W h^0_1$ should be taken into account along with the VBF process to improve the
overall signal efficiency. This is particularly true for $h^0_1$ masses below 
$\sim 90$ GeV where the Higgs--strahlung cross section exceeds that from VBF production.  
In another parton--level study \cite{cheung}, Higgs--strahlung  with  $h^0_1 \to
a^0_1 a^0_1 \to 4b$ was considered, and it was claimed that with a charged
lepton  for the $W$ boson and the fully tagged $4b$ final state  with Higgs mass
reconstruction, the signal could be disentangled from the background. 
Nevertheless, these results need to be  confirmed by MC and detector
simulations.\s

Note that in Ref.~\cite{lighthiggsteva2}, the $ h^0_1 \to 4\gamma$ final state
topology was advocated to be useful for the LHC at very high luminosity if the
branching ratio of the decay exceeds the level of $10^{-4}$. However, in
general, this decay has a smaller branching ratio and again, a detailed 
simulation which takes into account the experimental environment is lacking
here. The scope of other decays, such as $a^0_1a^0_1\to jjjj$ and
$jj\,\tau^+\tau^-$  where $j$ represents a light quark jet, is  expected to be
very much reduced, while the possibilities from $a^0_1a^0_1\to 4\tau$ are
currently being explored, in both vector boson  and Higgs--strahlung production
processes.\s

\underline{In scenarios with a very light scalar Higgs boson} such as our point
P4, constraints from Higgs searches at LEP do not allow for $h^0_1$ masses below
about 10 GeV, hence the main decay mode would be $h^0_1 \to b\bar b$ while  the
decay $h^0_1 \to \tau^+ \tau^-$ would have a branching ratio of the order of
7--8\%. The studies discussed above for very light pseudoscalar Higgs bosons but
with the SM--like Higgs boson being the $h^0_2$ state, which then decays into
two $h^0_1$ bosons, can therefore be adapted to this case. In particular, the
situation for point P4 would be very similar to that of point P1 as  the
production cross sections, the decay branching ratios and the masses of the
involved primary and secondary Higgs bosons are very similar.  \s

Ref.~\cite{lightHiggs} considered a particularly challenging NMSSM scenario,
with a Higgs spectrum very similar to that of the MSSM, i.e., nearly degenerate
(doublet dominated), heavy charged, scalar and pseudoscalar states and a light
scalar Higgs boson at around $120$--$140$~GeV, but including an additional
singlet-dominated scalar and a pseudoscalar; such a scenario is somewhat similar
to our point P4. Despite of having a reasonably large production cross-sections
at the LHC, this light Higgs boson would be difficult to observe since its main
hadronic decays cannot be easily disentangled from the QCD backgrounds.\s

In addition, some studies performed in the CP--violating MSSM, in which
the $h^0_1$ boson can also be very light with reduced couplings to gauge bosons, 
can also be adapted to this context. Note however, that in the case of the
CP--violating MSSM, one would also have a light $h^\pm$ boson, while in the 
NMSSM this state can be very heavy.  \s

Finally, \underline{in scenarios in which all NMSSM  Higgs bosons are relatively
light} as in point P5, no detailed analysis has been performed. To our
knowledge, the only study that is available is the one performed in
Ref.~\cite{egh1} in which the ATLAS and CMS signal significances for the MSSM,
assuming a high luminosity of 300 fb$^{-1}$, was rescaled to take into account
the reduced couplings of the various Higgs particles to gauge bosons and top
quarks. The effect of almost overlapping resonances, new decay modes or
production channels not present in the MSSM analyses of the ATLAS and CMS
collaborations have not been considered. In fact, the situation in this
scenario  looks similar to that of the intense coupling regime of the MSSM
\cite{ICR}, in which the three neutral Higgs bosons have  masses in the range
100--140 GeV. It was shown in the detailed simulation of Ref.~\cite{intense}
that while it would be very difficult to resolve all Higgs resonances, it would
be at least possible to detect one or two Higgs states. However, in the latter
case, the value of $\tan\beta$ is assumed to be large, leading to strongly
enhanced cross sections in the $gg$ fusion and $b\bar b$~Higgs processes, while
in the NMSSM one could have relatively moderate $\tb$ values as in P5 and thus,
smaller production rates. 

\subsubsection*{4.3 ATLAS strategy for NMSSM $h^0_1\rightarrow a^0_1a^0_1$ searches}

At the  ATLAS collaboration \cite{AtlasTDR}, current efforts to find suitable
search channels for special NMSSM phenomenological scenarios are being focused
on the vector boson fusion production of a scalar Higgs boson with relatively
low mass and subsequent decay via a pair of pseudoscalar Higgs bosons into four
$\tau$-leptons, $h^0_1 \rightarrow a^0_1a^0_1 \rightarrow 4\tau$. This decay
chain is favored in points  P2 and P3 proposed here. The main emphasis is
presently given to the case where all four $\tau$-leptons decay leptonically.\s

 A typical feature of the vector boson fusion production mode is the so-called
tagging jets that are  produced from the quarks that are scattered off the heavy
vector bosons and merge to give the Higgs boson. These jets typically have high
energies and lie in different hemispheres in the forward- and backward regions 
of the detector. Cutting on this signature is an important means to suppress
background processes. Since there is no color flow between the quarks in the
vector boson fusion process, jet production in the  central detector region is
suppressed. In contrast, central emission is favored in QCD interactions which 
constitute important background processes at the LHC \cite{Rainwater:1997dg}.
Experimentally, this can be exploited by vetoing on additional jets in this
region.\s

The decay products of the Higgs boson typically lie in the central detector
region. In general, leptons from the same pseudoscalar Higgs boson form pairs
that lie  close to each other in the detector, the separation being sensitive to
the pseudoscalar mass.  The invariant mass of the lepton pair has to be lower
than the pseudoscalar mass, and is thus much lower than the  $Z^0$-mass. For
background processes including $Z^0$-bosons, the photon interference therefore
needs to be considered. In the experimentally most simple case, all four
$\tau$-leptons decay to muons in the process $h^0_1\rightarrow a^0_1a^0_1
\rightarrow 4\tau \rightarrow 4\mu + 4\nu_\mu +4\nu_\tau$.  Since muons do not
deposit considerable energy in the detector,  muons from a very close pair can
be also classified as isolated.  Decay channels including electrons need more
consideration, as their energy is deposited in the calorimeters by
electromagnetic showering. The possibility of separating nearby electrons or
finding un-isolated muons inside an electromagnetic shower from an electron
needs careful study with a full ATLAS detector simulation. \s

The transverse momentum of the stable leptons in this channel is rather low,
since a large part of the energy  is carried away by the eight  neutrinos in the
final state. Therefore, not in all cases will all four leptons be identified. It
might  therefore prove favorable to require only three leptons to be found. For
triggering, two muons (electrons)  with $p_T>$10 GeV (20 GeV) or one muon
(electron) with $p_T$$>$20 GeV (25 GeV) are needed.  It might also be considered
to require a minimum transverse momentum for the remaining leptons to avoid
having a high number of lepton fakes. Another feature of this channel is the
large missing momentum from the eight neutrinos in the final state.\s 

In spite of the eight neutrinos in the final state and the fact that one lepton
might remain unrecognized,  it is still possible to reconstruct the mass of the
scalar Higgs boson with help of the collinear approximation\cite{Ellis:1987xu}
which is also used for mass reconstruction in the VBF, Higgs~$\rightarrow
\tau\tau$ channel \cite{Asai:2004ws}.  Since the pseudoscalar bosons and the
$\tau$-leptons from their decays obtain a large Lorentz  boost due to the large
mass and $p_T$ of the Higgs boson, their decay products are emitted roughly in
the  direction of the original pseudoscalar. Exploiting momentum conservation in
the transverse plane yields the 4-momentum vectors of the two pseudoscalars and
thus the invariant mass of the scalar Higgs boson. The performance of this
algorithm for the $h^0_1\rightarrow a^0_1a^0_1 \rightarrow 4\tau \rightarrow
4\mu +8\nu$ channel is currently under study at ATLAS.\s 

Possible background processes for this channel are $t\bar{t}$ production, 
vector boson production in association with bottom or top quarks and production
of vector boson pairs with additional light jets. Here, the leptons can come
from decays of the  heavy vector bosons or from decays of the bottom quarks. The
tagging jets of the vector boson fusion channel might be faked by untagged
$b$-jets or by light jets from other sources in the event. It should be noted
that the production of a vector boson pair in association with two light jets
contains diagrams that have a structure similar to the vector boson fusion
process, with two heavy bosons being  scattered off by the incoming quarks and
no color flow between the quark lines. Possible methods to  separate these
background processes from the signal and their performance are currently studied
by the ATLAS-collaboration.

\subsubsection*{4.4 Prospects for the CMS experiment}

The final state topology $h^0_{1} \to a^0_1 a^0_1\to \tau^+\tau^-\tau^+\tau^-$
where the $\approx$ 100 GeV $h^0_1$ state is produced in the VBF process and
the  pseudoscalar $a^0_{1}$ boson has a mass $2 m_{\tau} < M_{a^0_{1}} < 2
m_{b}$  (so that the $a^0_1 \to \tau^+\tau^-$ decay mode is dominant) is 
currently under investigation by the CMS collaboration for the  $\mu^{\pm}
\mu^{\pm} \tau _{\rm jet}^{\mp} \tau _{\rm jet}^{\mp}$  final state containing
two same sign muons and two $\tau$ jets.  The $\tau$ leptons from the decays of
the light $a^0_1$ state are  approximately collinear and the non-isolated,
di--muon high level trigger is needed to select the signal events. The standard
CMS di-muon trigger with the relaxed isolation has the di-muon threshold of 10
GeV on both muons for a luminosity ${\cal L}=2\times10^{33}$ cm$^{-2}$s$^{-1}$
\cite{CMStdr} and, since  the muons from the signal events are very soft, the
lower thresholds are needed.  For instance, with a 7 GeV threshold the
efficiency is increased approximately by a factor of two, but  the QCD
background rate is also  increased by the same factor \cite{CMS-trig} which is
not acceptable. In order to cope with the rate, the same-sign relaxed di-muon
trigger was recently introduced \cite{CMS-trig2}.  A PYTHIA simulation shows
that the rate of di-muons from $b \bar{b}$ production is reduced by a factor of
four when asking for the two same sign muons to have a threshold of 5 GeV. The
off-line selection strategy requires the presence of the two same sign,
non-isolated muons with one track in the cone around each muon direction, thus
selecting the one prong $\tau$ decays. \s

For a point with masses $M_{h^0_1}\sim100$ GeV and  $M_{a^0_1} \sim 5$ GeV,
one   applies the following basic event selection cuts at the generation level:
$i)$  two same sign muons with $p_{T}>$7 GeV and $|\eta|<$2.1 with one  track of
$p_{T}>$1 GeV in a cone 0.3 around each muon; $ii)$  opposite charge for the
muon and the track; $iii)$  two $\tau$ jets with $p_{T}>$10 GeV and
$|\eta|<$2.1; $iv)$  two jets with $p_{T}>$30 GeV and $|\eta|<$4.5. After these
cuts,  and using the SM VBF Higgs production cross section which is
approximately 5.4 pb for the considered $h^0_1$ mass,   the cross section of the
signal after all selections is around 2 fb, leading to $\approx$ 60 events at a
luminosity of  30 fb$^{-1}$. The dominant backgrounds with two non isolated
muons from the $b \bar{b}$ and $t \bar{t}$ production processes are under
evaluation.\s 

For heavier $a^0_1$ bosons as in scenarios P2 and P3 where $M_{a^0_1} \sim 10$
GeV, the ``non isolation" requirement  should be relaxed since two $\tau$
leptons from the $a^0_{1} \to \tau \tau$ decay are more separated and  one
should accept zero or one track in the cone (note that the relaxed di-muon
trigger accepts both isolated and non isolated muons). One needs, however, to 
consider the backgrounds with isolated muons  from $W$ and $Z^0$ decays, in
addition to $b \bar{b}$ and $t \bar{t}$.\s

The $h^0_1 \to a^0_1 a^0_1 \to \tau^+\tau^-\tau^+\tau^-$ decay with $h^0_1$ 
produced in Higgs--strahlung with leptonic decays of the $W$  bosons, which  can
give a very clean and almost background free signal, is also being considered by
the CMS collaboration. The leptons coming from  $W$ decays allow to trigger on
the events and the CMS single isolated lepton trigger can be used
\cite{CMS-trig}.   The  signal is unique as one has two collinear $\tau$ leptons
due to the large boost of $a^0_1$ bosons and, potentially, the most interesting
final state is when one of the $\tau$s decays hadronically and the other decays
into muons.\s

As for the selection criteria and the muon and $\tau$ identification,  for each
$\tau$-jet candidate there must be a muon in the $\tau$-jet cone.  The $\tau$
jet and the muon are required to be oppositely charged and  the events with two
identified $\tau$ jets with $E_{T}>$ 10 GeV and two muons  with $p_{T}>$ 7 GeV
are selected. Unlike in the VBF case, the muons are not required  to have the
same sign. Finally, the events with extra jets on top of  two $\tau$ jets are
rejected. A preliminary, full detector simulation and reconstruction analysis of
the signal events in a scenario with $M_{h^0_1} \sim 100$ GeV and $M_{a^0_1}
\sim 5$ GeV shows that after all selection cuts,  one expects $\approx$ 10
events for 30 fb$^{-1}$ data if the SM cross  section of $\sim$  2.6 pb and a
branching ratio BR($h^0_1\rightarrow a^0_1a^0_1 ) \simeq  0.9$ are assumed.  
The potential backgrounds are $t\bar t, tW$, QCD multijet, $W$+jets and $b\bar
b$ events. Preliminary tests with full detector  simulation show that the
requirement of two $\tau$ jets, both having an  oppositely  charged muon in the
jet cone, as well as jet veto suppresses the backgrounds very  efficiently.

\subsection*{5. Conclusions} 

The NMSSM is a very interesting supersymmetric extension  of the SM, as it
solves the notorious $\mu$ problem of the MSSM and it has less fine tuning. It
also leads to an interesting collider phenomenology in some cases, in particular
in the Higgs sector, which is extended to contain an additional CP--even and 
CP--odd state as compared to  the MSSM. Hence, the  searches for the NMSSM Higgs
bosons will be rather challenging at the LHC in scenarios in which some neutral
Higgs particles are very light, opening the possibility of dominant Higgs to
Higgs decays, or when all Higgs bosons are relatively light but have reduced
couplings to the electroweak gauge bosons and to the top quarks, compared to the
SM Higgs case. These scenarios require  much more detailed phenomenological 
studies and experimental simulations  to make sure that at least one Higgs 
particle of the NMSSM will be observed at the LHC. \s

In this note, we have proposed benchmark points in which these difficult
scenarios are realized in a semi--unified NMSSM which involves a rather limited
number of input parameters at the grand unification scale and which fullfils
all  present collider and cosmological constraints. In three of the five
benchmark scenarios   introduced here, the lightest CP--even Higgs boson decays
mainly in two very light pseudoscalar Higgs states which subsequently decay into
two $b$ quarks or $\tau$ leptons, leading to four  fermion final states; in a
fourth scenario, the next--to--lightest CP--even Higgs boson has a mass of $\sim
120$ GeV and couplings to fermions and gauge bosons that are SM--like  but it 
decays into pairs of the lightest Higgs state, which then decays into a $b\bar
b$ pair; in a last scenario, all neutral and charged Higgs particles are light
(with masses less than $\sim 180$ GeV) and have weaker couplings to $W/Z^0$
bosons than the SM Higgs particle.  We have analysed the Higgs and
supersymmetric particle spectra of these benchmark scenarios,  discussed the
various decay and production rates as well as other phenomenological
implications and attempted to set the basis for the search strategies to be
followed by the ATLAS and CMS collaborations in order to observe  at least one
of the neutral Higgs states in these scenarios. 

\subsubsection*{Addendum} After the completion of this paper,  discussions of
various possibilities of the observation of the cascade decay $h^0_1 \to a^0_1
a^0_1 \to ...$ at hadron colliders appeared in Refs.~\cite{papers2007}. 

\subsubsection*{Acknowledgments}  AD is grateful to the Leverhulme Trust
(London, UK) for partial financial support in the form a Visiting Professorship
to the NExT Institute (Southampton \& RAL, UK) and to the Alexander von--Humbold
Foundation (Bonn, Germany). SM acknowledges  financial support from The Royal
Society (London, UK).  We acknowledge  support from the FP7  RTN 
MRTN-CT-2006-035505 (HEPTOOLS),  the French ANR project PHYS@COL\&COS and the
Indo--French Center IFCPAR for the project number 3004-2. We also thank the
organisers of the 2007 Les Houches workshop on `Physics at TeV Colliders'
for the nice working framework.

\end{document}